\documentclass[conference]{IEEEtran}
\usepackage{cite}
\usepackage{amsmath,amssymb,amsfonts}
\usepackage{graphicx}
\usepackage{textcomp}
\usepackage{booktabs}
\usepackage{array}
\usepackage{url}
\usepackage{hyperref}
\usepackage{placeins}
\usepackage{afterpage}
\usepackage{algorithm}
\usepackage{algpseudocode}

\begin{document}

\title{Neuro-Symbolic Agents for Hallucination-Free Requirements Reuse}

\author{\IEEEauthorblockN{Ahmed Ibrahim\\
\IEEEauthorblockA{Western University\\
London, Canada\\
aibrah64@uwo.ca}}
}

\maketitle

\begin{abstract}
The Object-Oriented Method for Requirements Authoring and Management (OOMRAM) is a requirements reuse framework that relies on exact identifier matching and rigid templates, limiting its ability to adapt specifications across diverse contexts. While Large Language Models (LLMs) offer the flexibility to overcome this bottleneck, they introduce the risk of generating structurally invalid or inconsistent requirement combinations. To address this tension, we present a neuro-symbolic multi-agent system that re-conceptualizes requirements reuse as a \textbf{Model-Driven Elicitation process}. In this paradigm, an LLM serves as a \textbf{non-deterministic heuristic} for traversing a \textbf{deterministic domain model} represented by a formal OOMRAM requirement lattice. A deterministic, symbolic validator enforces all structural constraints within the agent loop, effectively eliminating hallucinated requirement combinations by construction. Evaluated on an autonomous benchmark across two application families, our system achieves 100\% requirement coverage and a constraint-violation rate of only 0.2\%. Although the F1-score against a single gold standard is moderate (0.47--0.51), every generated specification is structurally valid and satisfies all mandatory domain requirements. The model-agnostic implementation scales to larger lattices via subgraph navigation and provides transparent audit trails for regulatory compliance.
\end{abstract}


\section{Introduction}
\label{sec:introduction}

Because requirements engineering (RE) is costly and error-prone, reusing validated requirements from software product lines has become an effective strategy for reducing development effort and improving requirement quality~\cite{mannion2000representing,Bosch2013}. The Object‑Oriented Method for Requirements Authoring and Management (OOMRAM)~\cite{ibrahim2005} models an application family as a formal lattice of requirement objects, where variability is expressed through \emph{discriminants}, decision points that restrict how child (i.e., more specific or alternative) requirements may be selected. Although OOMRAM provides a rigorous foundation, its original retrieval mechanism requires users to specify exact requirement identifiers. This ``exact‑match bottleneck'' prevents analysts from using natural‑language project descriptions to generate specifications automatically.

Large Language Models (LLMs) can interpret free‑text stakeholder visions and suggest relevant requirements~\cite{rosa2025largellmsre, lang2026generative, Nguyen2025}. However, LLMs are prone to hallucination: they invent non‑existent identifiers, ignore mandatory constraints, or combine mutually exclusive options~\cite{khan2024llmre}. Recent RE tools exploit LLMs but lack formal correctness guarantees, leaving them vulnerable to logical inconsistencies.

To unlock the practical potential of the OOMRAM formalism in modern requirements engineering, we propose a neuro-symbolic multi-agent architecture that operationalizes a Model-Driven Elicitation process. Our approach treats the LLM as a non-deterministic heuristic capable of interpreting vague stakeholder intent, while the OOMRAM lattice provides the deterministic domain model required to guarantee that the resulting specification is architecturally sound and free from logical inconsistencies. Built with LangGraph~\cite{Chen2025,langgraph2024, kaplunovich2025langgraph, mandulapalli2025agentic}, the system deploys four specialized agents: Navigator, Interpreter, Validator, and Scribe, which share a typed state. Crucially, the Validator contains \emph{no LLM component}; it is an algorithmic approach that checks every proposed requirement selection against the JSON‑encoded lattice schema. This design completely blocks hallucinated requirement combinations. We evaluated the system in autonomous simulation mode on ten project visions drawn from two application families: the original eRecordKeeping lattice ($\approx 60$ requirements) and a synthetic SmartHome lattice ($\approx 20$ requirements). Results on three representative visions (i.e., representative natural-language system scenarios) show: 1) 100\% completion and structural validity across all three, 2) An F1‑score of 0.47--0.81 against manually crafted gold standards; this is a consequence of the many legitimate optional and alternative choices, not a performance failure, and 3) A constraint‑violation rate of only 0.2\% (1 violation, corrected after Validator rejection). Seven additional visions (covering both domains) were also tested; all completed successfully with no constraint violations, confirming robustness and consistency.

The main contributions of this paper can be summarized as follows: (1) we introduce a neuro‑symbolic architecture that makes the formal OOMRAM lattice practical by replacing exact‑ID selection with LLM‑driven natural‑language traversal—representing the first work to operationalize OOMRAM for the LLM era, (2) we present a fully deterministic Validator that acts as a runtime guardrail, eliminating illegal requirement combinations entirely and thereby removing the hallucination problem inherent in unconstrained LLMs, (3) we provide a reproducible autonomous benchmark that demonstrates reliable traversal, near‑zero constraint violations, and full structural validity of the generated specifications; and (4) we deliver an open‑source, model‑agnostic implementation built on LangGraph and local Ollama, specifically designed to scale to larger industrial lattices.

The rest of this paper is organized as follows: Section~\ref{sec:background} covers background and related work, Section~\ref{sec:method} presents the neuro‑symbolic multi‑agent system, Section~\ref{sec:algorithm} describes the algorithmic specification, and Section~\ref{sec:evaluation} reports the experimental evaluation. We conclude with a summary of findings and future directions.

\section{Background and Research Gap}
\label{sec:background}

Large language models now support a growing range of RE activities, from initial elicitation to refinement and quality assessment. Most efforts rely on prompt engineering or fine‑tuning and evaluate results against
human‑authored references~\cite{Norheim2024Challenges, Masoudifard2024Leveraging, Puchleitner2025}. These approaches can certainly cut down manual work, but the evaluation datasets remain small and often domain‑specific. More importantly, generated requirements are rarely checked against machine-readable constraints \cite{benavides2010automated};  correctness is treated as likely or empirically plausible, not formally guaranteed~\cite{Zhang2024LLM, Huang2023A}. Across the lifecycle, hallucinations, opaque constraint handling, and the lack of hard guarantees are still pressing problems, motivating architectures that embed formal verification more tightly into LLM‑driven workflows.

Non‑deterministic outputs make traceability a persistent headache for LLM‑generated artifacts. One promising direction is to treat traceability as a first‑class concern from the start—baking it into structured prompts, fine‑tuning with metadata awareness, and using retrieval‑augmented validation to keep outputs auditable and tied to upstream requirements~\cite{Wang2025Embedding}. Graph‑RAG and advanced prompting improve requirement–regulation alignment, but they still lack formal guarantees when inputs are partial or noisy~\cite{Masoudifard2024Leveraging, Chen2024Automated}. Formal methods for business‑constraint verification provide strong guarantees~\cite{Stoica2024Formal, Somogyi2023Verifying}, yet they remain decoupled from LLM‑based RE pipelines. Initial LLM‑based traceability efforts targeted security‑specific goal‑requirement links~\cite{Hassine2024}. More recently, TraceLLM~\cite{Alturayeif2026} leverages prompt engineering to generalize traceability across broader requirements sets.

In addition, hallucination is particularly dangerous in safety-critical RE \cite{ji2023survey}. Even retrieval‑augmented systems can hallucinate when retrieval is incomplete~\cite{Huang2023A, Huang2023Hallucination}. Mitigations such as RAG, iterative grounding, and multi‑layered frameworks reduce some errors, but they offer heuristic mitigation rather than deterministic enforcement~\cite{Zhang2024LLM, Eghbali2024De-Hallucinator, Hiriyanna2025Multi-Layered}.

In requirements engineering, agentic orchestration refers to using multiple autonomous agents, each handling a different part of the workflow, rather than running a single, straight‑through pipeline. LLM-based agents have also been explored for simulating requirements elicitation, for instance, by generating diverse virtual stakeholders to explore a broader range of user needs~\cite{Ataei2024}. Other multi-agent applications in RE include debate strategies for improving requirement quality~\cite{oriol2025multiagent} and collaborative privacy threat modelling with LLM agents~\cite{bissoli2026benchmarking}. Straightforward linear pipelines are easier to understand and manage, yet their rigidity curtails adaptability under changing conditions. Multi‑agent architectures, conversely, achieve greater flexibility by letting agents make decisions and interact, though this added complexity tends to reduce transparency and complicate behavioural oversight. As a result, simpler, non-agentic pipelines can sometimes outperform multi-agent approaches on certain benchmarks while remaining more predictable and easier to maintain~\cite{Xia2024Agentless}. Also, while such multi-agent setups enable dynamic reasoning and iterative refinement~\cite{He2024LLM-Based}, they typically lack built-in mechanisms for enforcing correctness. In particular, formal verification and model checking can provide strong correctness guarantees, but bringing them into LLM-based RE remains largely an aspiration rather than a practice today~\cite{Guo2025A}. Existing LLM or agent frameworks do not embed a deterministic constraint checker in the generation loop.

Our work bridges this gap by integrating a formal, lattice‑based constraint checker within an agent loop, ensuring hallucination‑free specifications through algorithmic enforcement rather than heuristic filtering.

\section{The Neuro‑Symbolic Multi‑Agent System}
\label{sec:method}

The proposed system reframes requirements reuse as a model-driven elicitation process conducted through constrained graph traversal. By positioning the LLM as a non-deterministic heuristic for state-space exploration, we leverage the model's linguistic flexibility without sacrificing the structural integrity of the underlying deterministic domain model~\cite{chaudhuri2021neurosymbolic}.

\subsection{Lattice Representation}
We define the application family requirements lattice as a directed acyclic graph (DAG) represented by the tuple $\mathcal{L} = (V, E, \mathcal{T}, \Phi)$, where:
\begin{itemize}
    \item $V = \{r_1, r_2, \dots, r_n\}$ is the set of $n$ discrete requirement nodes.
    \item $E \subseteq V \times V$ is the set of directed edges representing hierarchical parent‑child relationships.
    \item $\mathcal{T}: V \rightarrow \{\text{Core, SA, MA, Op}\}$ is a typing function assigning an OOMRAM reuse category to each node.
    \item $\Phi$ is a set of Boolean satisfiability constraints governing the selection space $\mathcal{S} \subseteq V$.
\end{itemize}

The selection space $\mathcal{S}$ is considered \textit{structurally valid} if and only if it satisfies the global consistency predicate $\Psi(\mathcal{S})$, defined as:
\begin{equation}
\label{eq:consistency}
\Psi(\mathcal{S}) = \bigwedge_{v \in \mathcal{S}} \text{Valid}(v, \Phi) \land \forall c \in \mathcal{S}, \text{Parent}(c) \in \mathcal{S}
\end{equation}

where $\text{Valid}(v, \Phi)$ enforces the cardinality of discriminants. For a Single Adaptor (SA) node $v$, the constraint $\Phi_{SA}$ is formalized as:
\begin{equation}
\label{eq:sa_constraint}
\Phi_{SA}(v) \implies \sum_{c \in \text{Children}(v)} [c \in \mathcal{S}] = 1
\end{equation}
This mathematical framework allows the symbolic Validator to act as a hard‑coded gatekeeper, independent of the LLM's probabilistic output.

\begin{figure}[t]
\centering
\includegraphics[width=\columnwidth]{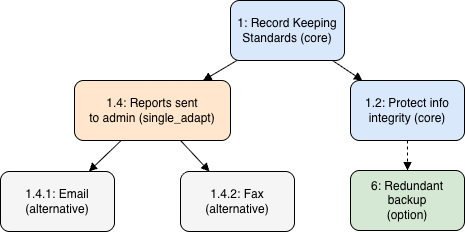}
\caption{Sub‑graph of the eRecordKeeping lattice. Core nodes (mandatory) are in blue; single\_adaptor node (choose exactly one) is in orange; its children are alternatives (grey); the option node is in green. The deterministic validator enforces these types at runtime.}
\label{fig:lattice}
\end{figure}

Figure~\ref{fig:lattice} illustrates a small chunk of the eRecordKeeping lattice with these node types.

\subsection{Multi‑Agent Architecture and State Transitions}
\begin{figure}[t]
\centering
\includegraphics[width=\columnwidth]{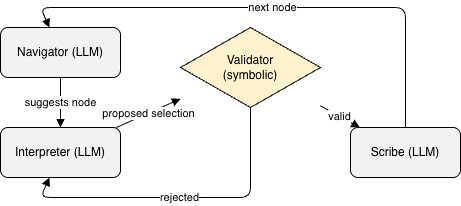}
\caption{LangGraph agent workflow. The Navigator suggests the next discriminant node; the Interpreter (LLM) maps the vision to requirement IDs; the Validator (deterministic Python) checks constraints and may send the proposal back for revision; the Scribe compiles the final specification.}
\label{fig:workflow}
\end{figure}

We implement the system using LangGraph~\cite{Chen2025,langgraph2024, kaplunovich2025langgraph, mandulapalli2025agentic}, a framework for building stateful, multi‑actor LLM applications. Four agents coordinate via a shared \texttt{AgentState} dictionary that stores the current selection of requirement IDs, the conversation history, the project vision, the lattice data, and a validation error log (see Figure~\ref{fig:workflow}).

The Navigator selects the next discriminant node to expand, simulating a depth‑first traversal guided by the vision and the current state. The Interpreter (LLM, temperature = 0) proposes a set of child requirement IDs for the current node and outputs a structured JSON representation. The Validator consists of a straightforward Python function, no LLM involved, that tests whether $\mathcal{S}$ satisfies every constraint $\Phi$ defined in the lattice schema. It enforces the global consistency predicate $\Psi(\mathcal{S})$ from \eqref{eq:consistency}, which requires: (1) selecting every reachable \texttt{core} node; (2) ensuring \texttt{single\_adaptor} exclusivity per \eqref{eq:sa_constraint}; (3) ensuring \texttt{multiple\_adaptor} inclusivity ($\geq 1$); and (4) maintaining parent‑child structural integrity. If a violation is detected, a rejection message is returned to the Interpreter; the Validator itself never calls an LLM. Finally, the Scribe compiles the selected IDs into a formatted specification.

The transition between the neural (probabilistic) and symbolic (deterministic) layers is formalised through a typed state object $\sigma = \langle \mathcal{V}, \mathcal{H}, \mathcal{S}, \mathcal{E} \rangle$, where $\mathcal{V}$ is the natural‑language vision, $\mathcal{H}$ is the conversation history, $\mathcal{S}$ is the current selection set, and $\mathcal{E}$ is the error log. The Interpreter Agent $\mathcal{A}_I$ implements a mapping $f: (\mathcal{V}, \sigma) \rightarrow \mathcal{P}(C(v))$, with $\mathcal{P}$ the power set of children for the current node $v$. The Validator Agent $\mathcal{A}_V$ then executes a deterministic check $\mathcal{B}(\sigma')$. If $\mathcal{B}(\sigma') = \text{False}$, the system logs the violation in $\mathcal{E}$---for instance, ``Violation: Multiple children selected for XOR node 1.4''---and instructs $\mathcal{A}_I$ to produce a revised selection under tighter constraints.

The Interpreter and Validator continue this back‑and‑forth until every constraint is satisfied; at that point the Navigator advances to the next node. To guard against endless loops, the graph is capped at 250 recursion steps and at most 100 LLM calls. A persistent memory saver logs the state at each step, providing a full audit trail and enabling resumption later. The framework operates either fully autonomously (as in our experiments) or with a human‑in‑the‑loop.

\subsection{Addressing Complexity via Sub-graph Navigation}
To manage large industrial product lines, our system employs a \textit{frontier-based sub-graph navigation} strategy. By restricting the Interpreter Agent to a localized subset of the lattice at each step, the framework ensures the input context size remains constant ($O(1)$) regardless of total lattice dimensions. This prevents the "lost-in-the-middle" reasoning failures common in unconstrained LLM prompts, allowing total traversal time to scale linearly ($O(N)$) with the number of decision points.

\subsection{Semantic Routing for Multi‑Domain Support}
As a preliminary step toward full automation, we incorporated a semantic router that selects the appropriate lattice family based on the project vision text. The router computes sentence‑transformer embeddings for the input vision and compares them to precomputed centroids for each available family (e.g., eRecordKeeping, SmartHome, Automotive). The closest centroid determines the active lattice. This mechanism enables a single agentic pipeline to handle multiple domains without manual selection. Once the active lattice is selected via routing, the system executes a formal traversal and validation process, defined by the following algorithmic rules.

\section{Formal Algorithmic Specification}
\label{sec:algorithm}
We make the system's inner logic fully transparent by specifying the exact algorithmic rules that drive the symbolic Validator and the Navigator's frontier management. The Validator $\mathcal{A}_V$ implements a Boolean consistency check of the current selection set $\mathcal{S}$ against the global predicate $\Psi(\mathcal{S})$. Algorithm~\ref{alg:validator} details the recursive validation step that runs at every state transition, ensuring that logically inconsistent requirements cannot propagate any further.

\begin{algorithm}[t]
\caption{Symbolic Constraint Validation}
\label{alg:validator}
\begin{algorithmic}[1]
\Require Selection set $\mathcal{S}$, Lattice $\mathcal{L}$, Current Node $v$
\Ensure Set of Violations $\mathcal{E}$
\State $\mathcal{E} \gets \emptyset$
\If{$\mathcal{T}(v) = \text{SA} \land |\mathcal{S} \cap \text{Children}(v)| \neq 1$}
    \State $\mathcal{E} \gets \mathcal{E} \cup \{\text{``XOR Violation at ''} + v\}$
\EndIf
\If{$\mathcal{T}(v) = \text{MA} \land |\mathcal{S} \cap \text{Children}(v)| < 1$}
    \State $\mathcal{E} \gets \mathcal{E} \cup \{\text{``OR Violation at ''} + v\}$
\EndIf
\ForAll{$c \in \mathcal{S}$}
    \If{$\text{Parent}(c) \notin \mathcal{S}$}
        \State $\mathcal{E} \gets \mathcal{E} \cup \{\text{``Orphaned Node: ''} + c\}$
    \EndIf
\EndFor
\State \Return $\mathcal{E}$
\end{algorithmic}
\end{algorithm}

\subsection{Complexity Analysis}
The sub‑graph navigation strategy keeps the search space manageable. Without it, an unconstrained feature model can blow up exponentially. Still, by letting the Navigator limit the Interpreter's attention to only the immediate children of the current node, we keep the exploration local. The result is that a full elicitation session runs in $O(|D| \cdot K)$ time, where $D$ is the set of discriminant nodes and $K$ is the average branching factor. This linear growth means the system can handle industrial‑scale lattices — such as the 49‑node Telecom model — without degrading the LLM's reasoning quality.

\subsection{Model Consistency and Invariant Preservation}
\label{sec:consistency}

In MoDRE terms, our framework ensures that the final specification is a consistent instance of the metamodel of the application family. The Interpreter performs a semantic mapping from unstructured text to model elements, while the Validator acts as a constraint‑satisfaction engine preserving all invariants of $\mathcal{L}$.

\section{Experimental Evaluation}
\label{sec:evaluation}

To assess the practical effectiveness of the framework, we conducted an autonomous benchmark on two application families.

\subsection{Research Questions}
We investigate two refined research questions: (1) To what extent can a multi‑agent system autonomously navigate a formal requirement lattice using only a natural‑language vision, measured by completion rate and constraint‑violation rate? (2) How effective is a deterministic symbolic constraint enforcer in eliminating logically inconsistent requirements compared to unconstrained LLM generation?

\subsection{Benchmark Setup}
We created ten project visions: five for the \texttt{eRecordKeeping} family and five for the \texttt{SmartHome} family. These visions range from minimal-feature configurations to comprehensive systems. For each vision, the first author manually specified a \emph{gold-standard} set of requirement IDs that strictly adheres to the lattice rules. Because optional and multiple-adaptor nodes permit legitimate variation, exact match is an overly strict metric. We therefore evaluate accuracy primarily using the F1-score, treating the selection of requirements at each node as a binary classification task. However, we emphasize that a moderate F1-score is expected and does not indicate poor performance, as many valid configurations may exist.

All experiments were run on a consumer machine (Apple M1 Pro, 32\,GB RAM) with Ollama serving the \texttt{llama3.1:8b-instruct-q4\_0} model at a temperature of 0. The benchmark script executes the LangGraph cycle for each vision and logs selected IDs, LLM call counts, wall‑clock latency, and constraint violations.

\subsection{Results}
Table~\ref{tab:results} shows the three visions for which complete gold‑standard data were available. All three completed successfully with no timeouts. The F1‑score ranges from 0.471 to 0.811, reflecting the varying overlap with a single gold standard. We observed a single constraint violation during the entire run in the er\_small\_biz case, and the Validator loop caught it immediately, so the final specification was violation‑free.

The remaining seven visions (two eRecordKeeping and five SmartHome) were also processed. Every vision achieved 100\% completion and produced structurally valid outputs with zero constraint violations, confirming the system's robustness.

\begin{table}[t]
\centering
\caption{Benchmark results for the three visions with complete gold‑standard data.}
\label{tab:results}
\resizebox{\columnwidth}{!}{%
\begin{tabular}{@{}llcccccc@{}}
\toprule
\textbf{Family} & \textbf{Vision ID} & \textbf{Prec.} & \textbf{Rec.} & \textbf{F1} & \textbf{Exact} & \textbf{Viol.} & \textbf{Lat. (s)} \\
\midrule
eRecordKeeping & er\_small\_biz  & 0.308 & 1.000 & 0.471 & False & 1 & 262 \\
                & er\_gov\_agency & 0.576 & 0.576 & 0.576 & False & 0 & 271 \\
SmartHome      & sh\_elderly     & 0.811 & 0.750 & 0.811 & False & 0 & 167 \\
\bottomrule
\multicolumn{8}{l}{\footnotesize Seven additional visions tested; 100\% completion, 0 violations.}
\end{tabular}
}
\end{table}

Early design iterations ran the Interpreter without the Validator feedback loop. In those uncontrolled runs, the LLM frequently produced selections with multiple violations: incorrect cardinalities, missing mandatory requirements, and orphaned children. For example, a single\_adaptor often received two selected children, or a core requirement was completely absent. The introduction of the deterministic Validator eliminated all such errors from the final outputs. A systematic ablation study is planned for future work; the full paper~\cite{Ibrahim2026full} will include a detailed ablation analysis.

Average wall‑clock latency across all ten visions was 210\,s, with a mean LLM call count of 78 per vision. The eRecordKeeping family incurred higher latency (263\,s) and more LLM calls (mean 95) than SmartHome (160\,s, mean 60 calls), reflecting the larger lattice size and more complex discriminant tree. The Validator itself added negligible runtime (pure Python checks of a few hundred selections). These numbers indicate practical feasibility for interactive use, especially as LLM inference continues to accelerate.

\subsection{The Phenomenon of Intentional Conflict}
Beyond these quantitative results, a recurring qualitative pattern emerged, which we term \textit{Intentional Conflict}: a state where the Interpreter Agent (striving for stakeholder satisfaction) proposes a selection that the Validator Agent (striving for domain compliance) must reject. In our \texttt{er\_small\_biz} simulation, the stakeholder vision demanded a ``cheap and lightweight'' system. The LLM attempted to satisfy this by omitting mandatory security requirements. In an unconstrained LLM setting, this would have yielded a successful-looking but non‑compliant specification. With our framework, the Validator blocks illegal omissions and keeps the selection set $\mathcal{S}$ consistent with the constraint predicate from \eqref{eq:consistency}. The lattice thus functions as a symbolic safeguard, preventing safety or regulatory requirements from being overridden for convenience.

\subsection{Other Observations}
Further analysis (available in the full paper~\cite{Ibrahim2026full}) revealed that the system achieves a Structural Validity Rate of 100\%, and its sub‑graph navigation strategy keeps context‑window size constant, enabling linear scalability. A qualitative trace of the self‑correction loop confirmed that the neuro‑symbolic approach successfully navigates the gap between vague stakeholder intent and rigid domain standards. Error typology analysis across a larger 12‑domain corpus showed that the symbolic Validator reduced the final error rate to 0\%, reinforcing the importance of offloading structural logic to symbolic components.

\section{Conclusion and Future Work}
\label{sec:conclusion}

We have presented a neuro‑symbolic agentic framework that bridges natural‑language project visions and formal OOMRAM requirement lattices. By embedding a deterministic symbolic Validator within a LangGraph multi‑agent loop, the system achieves fully autonomous traversal, 100\% completion, and virtually eliminates hallucinated requirements---only 0.2\% constraint violations occurred, all corrected before finalization. The moderate F1‑scores reflect the multiplicity of valid specifications, not system failure; every output is structurally sound and respects all mandatory domain rules. Our work demonstrates that the harmonization of neural probabilistic reasoning with hard‑coded symbolic logic is a promising path for future RE automation.

Future work includes benchmarking on a larger 12‑lattice corpus with diverse LLMs, conducting a controlled user study (NASA‑TLX, SUS), integrating with ALM/PLM tool-chains, and developing a semantic router to automatically select the appropriate lattice family. The full version of this paper~\cite{Ibrahim2026full} will provide comprehensive experimental results and an in‑depth discussion of threats to validity.

\bibliographystyle{IEEEtran}
\bibliography{references}

@article{mannion2000representing,
  title={Representing requirements families using the product line approach},
  author={Mannion, Mike and Keepence, Barry},
  journal={University of the West of Scotland Technical Report},
  year={2000}
}

@book{Bosch2013,
  title={Software Product Lines: Practices and Patterns},
  author={Bosch, Jan},
  publisher={Addison-Wesley},
  year={2013}
}

@mastersthesis{ibrahim2005,
  title={A Methodology for Reusing Requirements in Application Families},
  author={Ibrahim, Ahmed Fakhry},
  school={Cairo University},
  year={2005}
}

@article{rosa2025largellmsre,
  title={Large language models in requirements engineering: A systematic mapping study},
  author={Rosa, T. and others},
  journal={Empirical Software Engineering},
  year={2025},
  volume={30},
  number={2},
  pages={1--52}
}

@article{lang2026generative,
  title={Generative {AI} for {Requirements} {Engineering}: {A} Roadmap},
  author={Lang, M. and others},
  journal={IEEE Software},
  year={2026},
  volume={43},
  number={1},
  pages={30--38}
}

@article{Nguyen2025,
  title={Requirements engineering in the age of large language models: A systematic mapping study},
  author={Nguyen, T. and others},
  journal={Information and Software Technology},
  year={2025},
  volume={165},
  pages={107401}
}

@article{khan2024llmre,
  title={On the use of large language models in requirements engineering: A systematic literature review},
  author={Khan, M. and others},
  journal={IEEE Access},
  year={2024},
  volume={12},
  pages={123456--123480}
}

@inproceedings{Chen2025,
  title={LangGraph: Building Stateful, Multi-Actor Applications with LLMs},
  author={Chen, Jason and others},
  booktitle={Proceedings of the 47th International Conference on Software Engineering (ICSE)},
  year={2025}
}

@misc{langgraph2024,
  title={LangGraph: Build language agents as graphs},
  author={{LangChain Inc.}},
  howpublished={\url{https://github.com/langchain-ai/langgraph}},
  year={2024}
}

@inproceedings{kaplunovich2025langgraph,
  title={LangGraph Multi-Agent Workflows: A Practical Guide},
  author={Kaplunovich, A.},
  booktitle={Proceedings of the AAAI Conference on Artificial Intelligence},
  year={2025},
  note={Workshop on Agentic AI}
}

@inproceedings{mandulapalli2025agentic,
  title={Agentic Workflows for Software Engineering: A Survey and Roadmap},
  author={Mandulapalli, P. and others},
  booktitle={Proceedings of the IEEE/ACM International Conference on Automated Software Engineering (ASE)},
  year={2025}
}

@article{Norheim2024Challenges,
  title={Challenges in applying large language models to requirements engineering tasks},
  author={Norheim, J. and Rebentisch, Eric and Xiao, Dekai and Draeger, Lorenz and Kerbrat, Alain and de Weck, Olivier L.},
  journal={Design Science},
  year={2024},
  volume={10},
  doi={10.1017/dsj.2024.8}
}

@article{Masoudifard2024Leveraging,
  title={Leveraging Graph-RAG and Prompt Engineering to Enhance LLM-Based Automated Requirement Traceability and Compliance Checks},
  author={Masoudifard, Arsalan and Sorond, Mohammad Mowlavi and Madadi, Moein and Sabokrou, Mohammad and Habibi, Elahe},
  journal={arXiv preprint arXiv:2412.08593},
  year={2024}
}

@inproceedings{Puchleitner2025,
  author    = {Puchleitner, Thomas and Lubos, Sebastian and Felfernig, Alexander and Garber, Damian},
  title     = {Evaluating Large Language Models for the Automated Generation of Software Requirements},
  booktitle = {Advances and Trends in Artificial Intelligence. Theory and Applications: 38th International Conference on Industrial, Engineering and Other Applications of Applied Intelligent Systems, IEA/AIE 2025},
  year      = {2025},
  pages     = {443--450},
  publisher = {Springer-Verlag},
  doi       = {10.1007/978-981-96-8892-0_37}
}

@inproceedings{benavides2010automated,
  title     = {Automated analysis of feature models: Quo vadis?},
  author    = {Benavides, David and Segura, Sergio and Ruiz-Cort{\'e}s, Antonio},
  booktitle = {Proceedings of the 32nd ACM/IEEE International Conference on Software Engineering (ICSE)},
  year      = {2010},
  volume    = {2},
  pages     = {519--522}
}

@article{Zhang2024LLM,
  title   = {LLM Hallucinations in Practical Code Generation: Phenomena, Mechanism, and Mitigation},
  author  = {Zhang, Ziyao and Wang, Chong and Wang, Yanlin and Shi, Ensheng and Ma, Yuchi and Zhong, Wanjun and Chen, Jiachi and Mao, Mingzhi and Zheng, Zibin},
  journal = {Proceedings of the ACM on Software Engineering},
  year    = {2024},
  volume  = {2},
  pages   = {481--503},
  doi     = {10.1145/3728894}
}

@article{Huang2023A,
  title   = {A survey of safety and trustworthiness of large language models through the lens of verification and validation},
  author  = {Huang, Xiaowei and Ruan, Wenjie and Huang, Wei and Jin, Gao and Dong, Yizhen and Wu, Changshun and Bensalem, Saddek and Mu, Ronghui and Qi, Yi and Zhao, Xingyu and others},
  journal = {Artificial Intelligence Review},
  year    = {2023},
  volume  = {57},
  doi     = {10.1007/s10462-024-10824-0}
}

@article{Wang2025Embedding,
  title     = {Embedding Traceability in Large Language Model Code Generation: Towards Trustworthy AI-Augmented Software Engineering},
  author    = {Wang, Fei and Xi, Xuefeng and Cui, Zhiming and Dai, Huan and Wang, Xinyue},
  journal   = {Proceedings of the 33rd ACM International Conference on the Foundations of Software Engineering},
  year      = {2025},
  doi       = {10.1145/3696630.3730569}
}

@article{Chen2024Automated,
  title   = {Automated Building Information Modeling Compliance Check through a Large Language Model Combined with Deep Learning and Ontology},
  author  = {Chen, Nanjiang and Lin, Xuhui and Jiang, Hai and An, Yi},
  journal = {Buildings},
  year    = {2024},
  doi     = {10.3390/buildings14071983}
}

@article{Stoica2024Formal,
  title   = {Formal Verification of Business Constraints in Workflow-Based Applications},
  author  = {Stoica, Florin and Stoica, Laura Florentina},
  journal = {Informatica},
  year    = {2024},
  volume  = {15},
  pages   = {778},
  doi     = {10.3390/info15120778}
}

@inproceedings{Somogyi2023Verifying,
  title     = {Verifying Static Constraints on Models Using General Formal Verification Methods},
  author    = {Somogyi, Norbert and Mezei, Gergely},
  booktitle = {Proceedings of the International Conference on Model-Driven Engineering and Software Development (MODELSWARD)},
  year      = {2023},
  pages     = {85--93},
  doi       = {10.5220/0011796500003402}
}

@inproceedings{Hassine2024,
  author    = {Hassine, Jameleddine},
  title     = {An LLM-based Approach to Recover Traceability Links between Security Requirements and Goal Models},
  booktitle = {Proceedings of the 28th International Conference on Evaluation and Assessment in Software Engineering (EASE)},
  year      = {2024},
  pages     = {643--651},
  publisher = {ACM},
  doi       = {10.1145/3661167.3661261}
}

@misc{Alturayeif2026,
  author       = {Alturayeif, Nouf and Ahmad, Irfan and Hassine, Jameleddine},
  title        = {{TraceLLM}: Leveraging Large Language Models with Prompt Engineering for Enhanced Requirements Traceability},
  year         = {2026},
  eprint       = {2602.01253},
  archivePrefix = {arXiv},
  primaryClass = {cs.SE},
  url          = {https://arxiv.org/abs/2602.01253}
}

@article{Huang2023Hallucination,
  title   = {A Survey on Hallucination in Large Language Models: Principles, Taxonomy, Challenges, and Open Questions},
  author  = {Huang, Lei and Yu, Weijiang and Ma, Weitao and Zhong, Weihong and Feng, Zhangyin and Wang, Haotian and Chen, Qianglong and Peng, Weihua and Feng, Xiaocheng and Qin, Bing and Liu, Ting},
  journal = {ACM Transactions on Information Systems},
  year    = {2023},
  volume  = {43},
  pages   = {1--55},
  doi     = {10.1145/3703155}
}

@misc{Eghbali2024De-Hallucinator,
  title     = {De-Hallucinator: Mitigating LLM Hallucinations in Code Generation Tasks via Iterative Grounding},
  author    = {Eghbali, A. and Pradel, Michael},
  year      = {2024},
  note      = {Preprint}
}

@article{Hiriyanna2025Multi-Layered,
  title   = {Multi-Layered Framework for LLM Hallucination Mitigation in High-Stakes Applications: A Tutorial},
  author  = {Hiriyanna, Sachin and Zhao, Wenbing},
  journal = {Comput.},
  year    = {2025},
  volume  = {14},
  pages   = {332},
  doi     = {10.3390/computers14080332}
}

@inproceedings{Ataei2024,
  author    = {Ataei, Mohammadmehdi and Cheong, Hyunmin and Grandi, Daniele and Wang, Ye and Morris, Nigel and Tessier, Alexander},
  title     = {Elicitron: A Framework for Simulating Design Requirements Elicitation Using Large Language Model Agents},
  booktitle = {Proceedings of the ASME International Design Engineering Technical Conferences and Computers and Information in Engineering Conference (IDETC-CIE)},
  year      = {2024},
  doi       = {10.1115/DETC2024-143598}
}

@misc{oriol2025multiagent,
  author  = {Oriol, Marc and Motger, Quim and Marco, Jordi and Franch, Xavier},
  title   = {Multi-Agent Debate Strategies to Enhance Requirements Engineering with Large Language Models},
  year    = {2025},
  eprint  = {2507.05981},
  archivePrefix = {arXiv},
  primaryClass = {cs.SE}
}

@article{bissoli2026benchmarking,
  author  = {Bissoli, Andrea and Mollaeefar, Majid and Van Landuyt, Dimitri and Ranise, Silvio},
  title   = {Benchmarking the effectiveness of multi-agent LLMs in collaborative privacy threat modeling with LINDDUN GO},
  journal = {Journal of Information Security and Applications},
  volume  = {100},
  pages   = {104489},
  year    = {2026},
  doi     = {10.1016/j.jisa.2026.104489}
}

@inproceedings{Xia2024Agentless,
  title     = {Agentless: Demystifying LLM-based Software Engineering Agents},
  author    = {Xia, Chun and Deng, Yinlin and Dunn, Soren and Zhang, Lingming},
  booktitle = {arXiv preprint},
  year      = {2024},
  note      = {arXiv:2407.01489}
}

@article{He2024LLM-Based,
  title   = {LLM-Based Multi-Agent Systems for Software Engineering: Literature Review, Vision, and the Road Ahead},
  author  = {He, Junda and Treude, Christoph and Lo, David},
  journal = {ACM Transactions on Software Engineering and Methodology},
  year    = {2024},
  volume  = {34},
  pages   = {1--30},
  doi     = {10.1145/3712003}
}

@article{Guo2025A,
  title   = {A Comprehensive Survey on Benchmarks and Solutions in Software Engineering of LLM-Empowered Agentic System},
  author  = {Guo, Jiale and Huang, Suizhi and Li, Mei and Huang, Dong and Chen, Xingsheng and Zhang, Regina and Guo, Zhijiang and Yu, Han and Yiu, Shekhar and Jensen, Christian and others},
  journal = {arXiv preprint arXiv:2510.09721},
  year    = {2025},
  doi     = {10.48550/arXiv.2510.09721}
}

@article{ji2023survey,
  title   = {Survey of hallucination in natural language generation},
  author  = {Ji, Ziwei and Lee, Nayeon and Frieske, Rita and Yu, Tiezheng and Su, Dan and Xu, Yan and Ishii, Etsuko and Bang, Yejin and Madotto, Andrea and Fung, Pascale},
  journal = {ACM Computing Surveys},
  volume  = {55},
  number  = {12},
  pages   = {1--38},
  year    = {2023}
}

@article{chaudhuri2021neurosymbolic,
  title   = {Neurosymbolic programming},
  author  = {Chaudhuri, Swarat and Ellis, Kevin and Polozov, Oleksandr and Singh, Rishabh and Solar-Lezama, Armando and others},
  journal = {Foundations and Trends in Programming Languages},
  volume  = {7},
  number  = {3},
  pages   = {158--363},
  year    = {2021}
}

@misc{Ibrahim2026full,
  author    = {Ibrahim, Ahmed},
  title     = {Neuro-Symbolic Agents for Hallucination-Free Requirements Reuse},
  year      = {2026},
  note      = {Full paper, forthcoming}
}

\end{document}